\documentclass{iopart}
\usepackage{iopams}
\usepackage{revsymb}
\usepackage{graphicx}
\usepackage{cite}
\usepackage{bm}

\newcommand{\ket}[1]{|{#1}\rangle}
\newcommand{\bra}[1]{\langle{#1}|}
\newcommand{\bras}[2]{{}_{#2}\hspace*{-0.2mm}\langle{#1}|}
\newcommand{\ketbras}[3]{\ket{#1}_{#3}\hspace*{-0.2mm}\bra{#2}}
\newcommand{\brackets}[3]{{}_{#3}\hspace*{-0.2mm}\langle{#1}|{#2}\rangle_{#3}}

\begin{document}
\title{Entanglement purification through Zeno-like measurements}
\author{\textmd{K. YUASA, H. NAKAZATO and M. UNOKI}}
\address{Department of Physics, Waseda University, Tokyo %
169--8555, Japan}
\ead{yuasa@hep.phys.waseda.ac.jp}
\begin{abstract}
We present a novel method to purify quantum states,
i.e.~\textit{puri\-fi\-ca\-tion through Zeno-like measurements},
and show an application to \textit{en\-tangle\-ment
puri\-fi\-ca\-tion}.
\end{abstract}
\pacs{03.65.Xp, 03.67.Mn}
\section{Introduction}
One of the main obstacles to the experimental realisation of the
ideas of quantum computation and quantum information is
`decoherence'~\cite{ref:QInfoCompNielsen,ref:QInfoCompZeilinger}.
There are many attempts to overcome this problem, and several
approaches are proposed for it.
Among them are the `purification/distillation'
technologies~\cite{ref:QInfoCompZeilinger,ref:Purification},
i.e.~methods to extract pure states, especially entanglements,
from general mixed states recovering/preparing quantum coherence.
Contributing to this subject, we have recently proposed a novel
method to purify quantum states, i.e.~purification through
Zeno-like measurements~\cite{ref:qpf}.
Here we show that it is possible to apply it for extracting
\textit{entanglement}, which is the key resource to the
fundamentals of quantum computation and information.

\section{Purification through Zeno-like measurements}
\label{sec:Framework}
Let us recapitulate the framework of our
purification~\cite{ref:qpf}.
We consider two quantum systems X and A interacting with each
other.
The total system is initially in a general \textit{mixed} state
$\varrho_\mathrm{tot}$, from which we try to extract a pure state
in A\@.
We first make a measurement on X \textit{to confirm that it is in
a given state $\ket{\phi}_\mathrm{X}$}.
If it is found in the state $\ket{\phi}_\mathrm{X}$, the state of
the total system is projected by the projection operator
\begin{equation}
\mathcal{O}
=\ketbras{\phi}{\phi}{\mathrm{X}}\otimes\openone_\mathrm{A}.
\end{equation}
We then let the total system start to evolve under the
Hamiltonian $H_\mathrm{tot}$ and repeat the same measurement on X
at time intervals $\tau$.
After $N$ repetitions of successful confirmations, the state of
the total system, $\varrho_\mathrm{tot}^{(\tau)}(N)$, is given by
\begin{eqnarray}
\fl
\varrho_\mathrm{tot}^{(\tau)}(N)
&=&(\mathcal{O}e^{-iH_\mathrm{tot}\tau}\mathcal{O})^N
\varrho_\mathrm{tot}
(\mathcal{O}e^{iH_\mathrm{tot}\tau}\mathcal{O})^N/P^{(\tau)}(N)
=\ketbras{\phi}{\phi}{\mathrm{X}}\otimes
\varrho_\mathrm{A}^{(\tau)}(N),
\label{eqn:StateTotal}\\
\fl
\varrho_\mathrm{A}^{(\tau)}(N)
&=&\bm{(}V_\phi(\tau)\bm{)}^N\varrho_\mathrm{A}
\bm{(}V_\phi^\dag(\tau)\bm{)}^N/P^{(\tau)}(N),
\label{eqn:StateA}
\end{eqnarray}
where $\varrho_\mathrm{A}=\bras{\phi}{\mathrm{X}}
\varrho_\mathrm{tot}\ket{\phi}_\mathrm{X}$ is the state of A
after the zeroth confirmation,
\begin{equation}
V_\phi(\tau)
\equiv\bras{\phi}{\mathrm{X}}e^{-iH_\mathrm{tot}\tau}
\ket{\phi}_\mathrm{X}
\end{equation}
is a projected time-evolution operator, and $P^{(\tau)}(N)$ is
the normalisation factor,
\begin{equation}
\fl
P^{(\tau)}(N)
=\Tr[(\mathcal{O}e^{-iH_\mathrm{tot}\tau}\mathcal{O})^N
\varrho_\mathrm{tot}(\mathcal{O}e^{iH_\mathrm{tot}\tau}
\mathcal{O})^N]
=\Tr_\mathrm{A}[\bm{(}V_\phi(\tau)\bm{)}^N\varrho_\mathrm{A}
\bm{(}V_\phi^\dag(\tau)\bm{)}^N].
\label{eqn:Yield}
\end{equation}
Note that we retain only the events where X is found in the state
$\ket{\phi}_\mathrm{X}$ at \textit{every} measurement; other
events, resulting in failure to purify A, are discarded.
The normalisation factor $P^{(\tau)}(N)$ in~(\ref{eqn:Yield}) is
nothing but the probability for the \textit{successful events}
and is the probability of obtaining the state given
in~(\ref{eqn:StateTotal}) and~(\ref{eqn:StateA}).

Let us assume for simplicity that the eigenvalues $\lambda_n$ of
$V_\phi(\tau)$, which are complex-valued in general, are discrete
and $V_\phi(\tau)$ is decomposed (diagonalised) as
\begin{equation}
V_\phi(\tau)=\sum_n\lambda_n\ketbras{u_n}{v_n}{\mathrm{A}},
\label{eqn:SpectralDecomp}
\end{equation}
where $\ket{u_n}_\mathrm{A}$ and $\bras{v_n}{\mathrm{A}}$ are the
right and left eigenvectors of $V_\phi(\tau)$, respectively,
belonging to the eigenvalue $\lambda_n$ and satisfy
$\brackets{v_n}{u_m}{\mathrm{A}}=\delta_{mn}$.
(The right eigenvectors are normalised as
$\brackets{u_n}{u_n}{\mathrm{A}}=1$ in the following.)
It is now easy to observe the asymptotic behaviour of the state
of A in~(\ref{eqn:StateA}).
Since the eigenvalues $\lambda_n$ are bounded as $0
\le|\lambda_n|\le1$, each term in the expansion of
$\bm{(}V_\phi(\tau)\bm{)}^N$ decays away and a single term
dominates asymptotically as the number of measurements, $N$,
increases,
\begin{equation}
\bm{(}V_\phi(\tau)\bm{)}^N
=\sum_n\lambda_n^N\ketbras{u_n}{v_n}{\mathrm{A}}
\to\lambda_0^N\ketbras{u_0}{v_0}{\mathrm{A}}
\quad\mbox{as $N$ increases},
\label{eqn:Meachanism}
\end{equation}
\textit{provided the largest (in magnitude) eigenvalue
$\lambda_0$ is unique, discrete and non\-degenerate}.
Then, the state of A in~(\ref{eqn:StateA}) accordingly approaches
the pure state $\ket{u_0}_\mathrm{A}$,
\begin{equation}
\varrho_\mathrm{A}^{(\tau)}(N)\to\ketbras{u_0}{u_0}{\mathrm{A}}
\quad\mbox{as $N$ increases}.
\label{eqn:Purification}
\end{equation}
This is the purification we have found recently~\cite{ref:qpf}:
extraction of a pure state $\ket{u_0}_\mathrm{A}$ from an
\textit{arbitrary} mixed state $\varrho_\mathrm{A}$ through
repeated measurements on X\@.
Since we repeat measurements (on X) as in the case of the quantum
Zeno effect~\cite{ref:QZE}, we call such measurements `Zeno-like
measurements'.\footnote{It should be noted however that the time
interval $\tau$ in this scheme is not necessarily small as in the
ordinary Zeno measurements, and the
purification~(\ref{eqn:Purification}) is not due to the quantum
Zeno effect.}

The above observation shows that the assumption of the `spectral
decomposition'~(\ref{eqn:SpectralDecomp}) is not essential but
the existence of the \textit{unique, discrete and nondegenerate
largest eigenvalue $\lambda_0$} is crucial for the purification.
Furthermore, note the asymptotic behaviour of the `success
probability' $P^{(\tau)}(N)$: it decays asymptotically as
\begin{equation}
P^{(\tau)}(N)
\to|\lambda_0|^{2N}\bras{v_0}{\mathrm{A}}\varrho_\mathrm{A}
\ket{v_0}_\mathrm{A}
\quad\mbox{as $N$ increases},
\label{eqn:Decay}
\end{equation}
which is dominated by the eigenvalue $\lambda_0$.
Efficient purification is possible if it satisfies the condition
$|\lambda_0|=1$, which suppresses the decay in~(\ref{eqn:Decay}).
At the same time, if the other eigenvalues are much smaller than
$\lambda_0$ in magnitude, purification is achieved faster.
Hence,
\begin{equation}
|\lambda_0|=1,\qquad
|\lambda_n/\lambda_0|\ll1\quad\mathrm{for}\quad n\neq0
\label{eqn:Optimization}
\end{equation}
are the conditions for the \textit{efficient purification}, which
we try to achieve by adjusting parameters such as $\tau$,
$\ket{\phi}_\mathrm{X}$ and those in the Hamiltonian
$H_\mathrm{tot}$.

\section{Application to entanglement purification}
Now we show an interesting application of the above scheme:
application to \textit{en\-tangle\-ment puri\-fi\-ca\-tion}.
Let us consider, for example, a three-qubit system with the
Hamiltonian
\begin{equation}
\fl
H_\mathrm{tot}
=\Omega\frac{1+\sigma_3^\mathrm{X}}{2}
+\Omega\frac{1+\sigma_3^\mathrm{A}}{2}
+\Omega\frac{1+\sigma_3^\mathrm{B}}{2}
+g(\sigma_+^\mathrm{X}\sigma_-^\mathrm{A}
+\sigma_-^\mathrm{X}\sigma_+^\mathrm{A})
+g(\sigma_+^\mathrm{X}\sigma_-^\mathrm{B}
+\sigma_-^\mathrm{X}\sigma_+^\mathrm{B}),
\label{eqn:ThreeQubitSystem}
\end{equation}
where $\sigma_i\,(i=1,2,3)$ are the Pauli operators and
$\sigma_\pm=(\sigma_1\pm i\sigma_2)/2$ are the ladder operators.
Qubit X is coupled to qubits A and B\@.
We confirm X to be in the state $\ket{\phi}_\mathrm{X}$
repeatedly at time intervals $\tau$ and end up with an extraction
of an entanglement between A and B, which are initially in a
general mixed state $\varrho_\mathrm{tot}$.

The projected time-evolution operator $V_\phi(\tau)$ is given, in
this case, by
\begin{eqnarray}
\fl
V_\phi(\tau)
&=&\ketbras{\Psi^-}{\Psi^-}{\mathrm{AB}}
\,\bigl(
|\beta|^2+|\alpha|^2e^{-i\Omega\tau}
\bigr)\,e^{-i\Omega\tau}\nonumber\\
\fl
&&{}+\ketbras{\downarrow\downarrow}{\downarrow\downarrow}%
{\mathrm{AB}}\,\bigl(
|\beta|^2+|\alpha|^2e^{-i\Omega\tau}\cos\sqrt{2}g\tau
\bigr)\nonumber\\
\fl
&&{}+\ketbras{\Psi^+}{\Psi^+}{\mathrm{AB}}
\,\bigl(
|\beta|^2+|\alpha|^2e^{-i\Omega\tau}
\bigr)\,e^{-i\Omega\tau}\cos\sqrt{2}g\tau\nonumber\\
\fl
&&{}+\ketbras{\uparrow\uparrow}{\uparrow\uparrow}{\mathrm{AB}}
\,\bigl(
|\beta|^2\cos\sqrt{2}g\tau
+|\alpha|^2e^{-i\Omega\tau}
\bigr)\,e^{-i2\Omega\tau}\nonumber\\
\fl
&&{}-i\,\bigl(
\alpha^*\beta\ketbras{\downarrow\downarrow}{\Psi^+}{\mathrm{AB}}
+\alpha\beta^*\ketbras{\Psi^+}{\downarrow\downarrow}{\mathrm{AB}}
\bigr)\,
e^{-i\Omega\tau}\sin\sqrt{2}g\tau\nonumber\\
\fl
&&{}-i\,\bigl(
\alpha\beta^*\ketbras{\uparrow\uparrow}{\Psi^+}{\mathrm{AB}}
+\alpha^*\beta\ketbras{\Psi^+}{\uparrow\uparrow}{\mathrm{AB}}
\bigr)\,
e^{-i2\Omega\tau}\sin\sqrt{2}g\tau,
\end{eqnarray}
where $\ket{\Psi^\pm}_\mathrm{AB}
=(\ket{\uparrow\downarrow}_\mathrm{AB}
\pm\ket{\downarrow\uparrow}_\mathrm{AB})/\sqrt{2}$ and
$\ket{\phi}_\mathrm{X}=\alpha\ket{\uparrow}_\mathrm{X}
+\beta\ket{\downarrow}_\mathrm{X}$,
with $\ket{{\uparrow}({\downarrow})}$ the eigenstates of
$\sigma_3$ belonging to the eigenvalues $+1\,(-1)$.
Since the Hamiltonian~(\ref{eqn:ThreeQubitSystem}) is symmetric
under the exchange between A and B, $V_\phi(\tau)$ splits into
the singlet and triplet sectors, and the singlet state
$\ket{\Psi^-}_\mathrm{AB}$ is one of the four eigenstates of
$V_\phi(\tau)$.
It is hence possible to extract a Bell state
$\ket{\Psi^-}_\mathrm{AB}$ provided (i)~the corresponding
eigenvalue
\begin{equation}
\lambda_{\Psi^-}=e^{-i\Omega\tau}\,\bigl(
|\beta|^2+|\alpha|^2e^{-i\Omega\tau}
\bigr)
\end{equation}
is larger (in magnitude) than any other eigenvalue, and the
extraction is efficient if the conditions
(\ref{eqn:Optimization}), i.e.~(ii)~$|\lambda_{\Psi^-}|=1$ and
(iii)~$|\lambda_n/\lambda_{\Psi^-}|\ll1\,(n\neq\Psi^-)$, are
fulfilled.

Condition~(ii) is achieved by tuning $\tau$ such as
\begin{equation}
|\Omega|\tau=2n\pi\quad(n=0,1,2,\ldots),
\label{eqn:Tuning}
\end{equation}
and condition~(i) is possible unless $\ket{\phi}_\mathrm{X}
=\ket{\uparrow}_\mathrm{X}$ or $\ket{\downarrow}_\mathrm{X}$.
Let us hence consider, under the condition~(\ref{eqn:Tuning}),
the case where
\begin{equation}
\alpha=\beta=1/\sqrt{2},\quad\mbox{i.e.}\quad
\ket{\phi}_\mathrm{X}
=\ket{\rightarrow}_\mathrm{X}
\equiv\bigl(
\ket{\uparrow}_\mathrm{X}+\ket{\downarrow}_\mathrm{X}
\bigr)/\sqrt{2}.
\label{eqn:ConfirmX}
\end{equation}
In this case, three other eigenvalues than $\lambda_{\Psi^-}$ are
given by
\begin{equation}
\fl
\lambda_{\Phi^-}=\cos^2\!\frac{g\tau}{\sqrt{2}},\quad
\lambda_\pm=1-\frac{1}{2}\sin\frac{g\tau}{\sqrt{2}}\left(
3\sin\frac{g\tau}{\sqrt{2}}
\pm\sqrt{1-9\cos^2\!\frac{g\tau}{\sqrt{2}}}
\right),
\end{equation}
whose magnitudes are less than $1$, fulfilling condition~(i),
when
\begin{equation}
|g|\tau/\sqrt{2}\neq m\pi/2\quad(m=0,1,2,\ldots).
\label{eqn:Coupling}
\end{equation}
Therefore, the set of such parameters as~(\ref{eqn:Tuning}),
(\ref{eqn:ConfirmX}) and (\ref{eqn:Coupling}) enables us to
extract the Bell state $\ket{\Psi^-}_\mathrm{AB}$ from a general
mixed state $\varrho_\mathrm{tot}$.

\begin{figure}
\begin{center}
\includegraphics[width=0.6\textwidth]{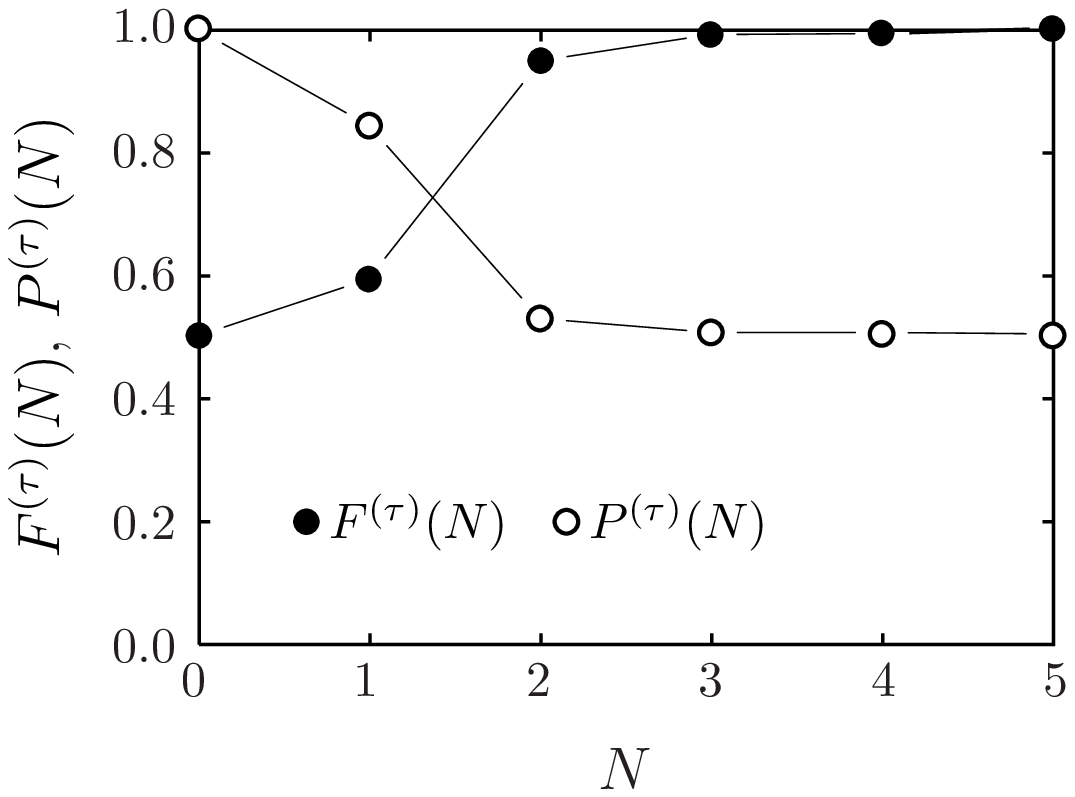}
\end{center}
\caption{Extraction of a Bell state $\ket{\Psi^-}_\mathrm{AB}$
from a product state $\varrho_\mathrm{tot}=\ketbras{\rightarrow}%
{\rightarrow}{\mathrm{X}}\otimes\ketbras{\uparrow}{\uparrow}%
{\mathrm{A}}\otimes\ketbras{\downarrow}{\downarrow}{\mathrm{B}}$
[or from a mixed state $\varrho_\mathrm{tot}
=\ketbras{\rightarrow}{\rightarrow}{\mathrm{X}}
\otimes(\ketbras{\uparrow\downarrow}{\uparrow\downarrow}%
{\mathrm{AB}}+\ketbras{\downarrow\uparrow}{\downarrow\uparrow}%
{\mathrm{AB}})/2$].
Parameters are $\tau=2\pi/\Omega$, $\ket{\phi}_\mathrm{X}
=\ket{\rightarrow}_\mathrm{X}$ and $g=0.25\Omega$ ($\Omega>0$).}
\label{fig:Fidelity}
\end{figure}

The extraction of the Bell state $\ket{\Psi^-}_\mathrm{AB}$ is
demonstrated in figure~\ref{fig:Fidelity}, where the fidelity to
the target state $\ket{\Psi^-}_\mathrm{AB}$, defined by
$F^{(\tau)}(N)\equiv\bras{\Psi^-}{\mathrm{AB}}
\varrho_\mathrm{AB}^{(\tau)}(N)\ket{\Psi^-}_\mathrm{AB}$, and the
success probability $P^{(\tau)}(N)$ versus the number of
measurements, $N$, are shown for the initial state
$\varrho_\mathrm{tot}=\ketbras{\rightarrow}{\rightarrow}%
{\mathrm{X}}\otimes\ketbras{\uparrow}{\uparrow}{\mathrm{A}}
\otimes\ketbras{\downarrow}{\downarrow}{\mathrm{B}}$, a (pure
but) product state [or for a mixed state $\varrho_\mathrm{tot}
=\ketbras{\rightarrow}{\rightarrow}{\mathrm{X}}
\otimes(\ketbras{\uparrow\downarrow}{\uparrow\downarrow}%
{\mathrm{AB}}+\ketbras{\downarrow\uparrow}{\downarrow\uparrow}%
{\mathrm{AB}})/2$].
It is clear that the Bell state $\ket{\Psi^-}_\mathrm{AB}$ is
extracted after only $4$ or $5$ measurements (in this case,
$|\lambda_{\Phi^-}|\simeq0.20$ and $|\lambda_\pm|\simeq0.44$).
Since condition~(ii) is fulfilled, the decay of the success
probability $P^{(\tau)}(N)$ is suppressed, yielding
$P^{(\tau)}(N)\to\bras{\Psi^-}{\mathrm{AB}}\varrho_\mathrm{AB}
\ket{\Psi^-}_\mathrm{AB}$\,($=1/2$ in figure~\ref{fig:Fidelity}),
which means that the $\ket{\Psi^-}_{\mathrm{AB}}$ component
contained in the initial state $\varrho_\mathrm{AB}$ (after the
zeroth measurement on $\varrho_\mathrm{tot}$) is fully extracted.
In this sense, the extraction is \textit{optimal}.

\section{Concluding remarks}
The above example clearly and explicitly shows that our
purification scheme works for entanglement purification.
It is quite simple: one has simply to repeat one and the same
measurement.
At the same time, the purification can be made optimal: the
optimal success probability is attainable as in the above
example.
Since the basic framework presented in
section~\ref{sec:Framework} is general, it possesses wide
potential applicabilities in various settings for quantum
computation and quantum information.

\section*{Acknowledgements}
This work is partly supported by Grants-in-Aid for Scientific
Research (C) from the Japan Society for the Promotion of Science
(No.~14540280) and Priority Areas Research (B) from the Ministry
of Education, Culture, Sports, Science and Technology, Japan
(No.~13135221), by a Waseda University Grant for Special Research
Projects (No.~2002A--567), and by the bilateral Italian--Japanese
project 15C1 on `Quantum Information and Computation' of the
Italian Ministry for Foreign Affairs.

\end{document}